\documentclass[aps,twocolumn]{revtex4}
\usepackage{graphicx,xcolor}
\usepackage{amsmath}
\usepackage{hyperref}
\newcommand{\fsa}[1]{\langle{#1}\rangle}

\newcommand{\jl}[1]{\textcolor{black}{#1}}
\begin{document}

\title{Control of the bootstrap current in approximately quasi-axisymmetric magnetic fields}

\author{J.L. Velasco}
\affiliation{Laboratorio Nacional de Fusi\'on, CIEMAT, 28040 Madrid, Spain}

\author{I. Calvo}
\affiliation{Laboratorio Nacional de Fusi\'on, CIEMAT, 28040 Madrid, Spain}

\author{J. M. Garc\'ia-Rega\~na}
\affiliation{Laboratorio Nacional de Fusi\'on, CIEMAT, 28040 Madrid, Spain}

\date{\today}

\begin{abstract}

Quasi-axisymmetric stellarators are the stellarator analogue of axisymmetric tokamaks, retaining many of their favorable confinement properties, their {compactness} and their relative coil simplicity, while avoiding their principal limitation, the need for an inductively driven plasma current. Despite these attractive physics properties, the development of quasi-axisymmetric configurations has been severely constrained by the absence of an experimentally validated divertor concept compatible with their large bootstrap current. In this Letter, approximately quasi-axisymmetric fields, complemented with piecewise omnigenous perturbations, are proposed as the basis for a new strategy towards a stellarator reactor that {potentially achieves, simultaneously,} simple coil geometries, tokamak-like confinement and, through tailoring of the bootstrap current, compatibility with an island divertor. Implications for attaining a high bootstrap current fraction in tokamak devices are also discussed.

\end{abstract}

\maketitle

\section{Introduction}\label{SEC_INTRO}

Magnetic confinement fusion aims to confine a hot plasma by means of strong magnetic fields that form nested toroidal flux surfaces. One of the central challenges is to limit transport processes across flux surfaces so that the plasma achieves the temperatures and densities required for fusion reactions to proceed efficiently~\cite{lawson1957criteria}.

The tokamak~\cite{arcimovich1968tokamak} has been the leading magnetic confinement concept because it combines a relatively simple geometry with excellent confinement properties. Its axisymmetry results in favorable neoclassical transport (the one associated with particle trajectories in an inhomogeneous magnetic field and collisions), and decades of experimental development have led to a relatively {well-established} route toward reactor-scale devices. However, despite these strengths, the tokamak suffers from intrinsic limitations. Its operation relies on a large, mainly inductive, plasma current, which enforces pulsed operation and makes the device vulnerable to current-driven instabilities and potentially disruptive events. This poses significant challenges for steady-state reactor operation. Strategies to mitigate these weaknesses include non-inductive current drive (e.g.~\cite{freethy2024step}) and operation with a high bootstrap current fraction \cite{goldston1994advanced,ding2022hbpol}. Nevertheless, current drive displays degraded efficiency in a reactor-size device, and the bootstrap current, a self-generated current caused by neoclassical effects, is limited by the size of the plasma gradients. Stellarators~\cite{spitzer1958stellarator} overcome these limitations by generating the confining magnetic field entirely through external coils. This facilitates steady-state operation and eliminates disruptions, but requires fully three-dimensional magnetic fields whose geometry must be carefully optimized to achieve acceptable transport~\cite{helander2012stell}. 

In both tokamaks and stellarators, charged particles approximately follow the magnetic field lines, executing fast gyromotion around them. On time scales much longer than the gyroperiod, magnetic field inhomogeneities and curvature give rise to \textit{drifts}, causing particles to slowly separate from individual field lines and, in particular, to move across flux surfaces. If the ratio between the particle energy $\mathcal{E}$ and the magnetic moment $\mu$ is larger than the maximum value of the magnetic field strength $B=|\mathbf{B}|$ on the flux surface, particles are \textit{passing}: their velocity component parallel to the magnetic field never vanishes, they explore the whole flux surface and remain, on average, tied to it. Particles with smaller $\mathcal{E}/\mu$ are \textit{trapped}: they bounce back and forth between \textit{bounce points}, where the parallel velocity vanishes, typically tracing banana-shaped trajectories in a small region of the flux surface. In axisymmetric tokamaks, the radial drift of these particles averages to zero; this {contrasts with unoptimized stellarators, where secular radial drifts lead to} unacceptably large transport across flux surfaces.

All strategies for the optimization of stellarators with respect to neoclassical transport can be understood in terms of the second adiabatic invariant 
\begin{equation}
J \equiv 2\int_{l_{b_1}}^{l_{b_2}}\mathrm{d}l \sqrt{2\left(\mathcal{E}-\mu B\right)}\,,\label{EQ_J}
\end{equation}
defined for trapped particles. Here, $l$ is the arc-length {along the field line}, and $l_{b_1}$ and $l_{b_2}$ the bounce-point positions, the points where $\mathcal{E}/\mu=B$. Ensuring that the contours of constant $J$ are aligned with the flux surfaces, as in omnigenous fields~\cite{hall1975omni}, implies that the (orbit-averaged) radial drift of trapped particles is zero. As a consequence, in omnigenous fields, low collisionality transport is described by neoclassical coefficients proportional to the collisionality, as in a tokamak, rather than exhibiting the $1/\nu$ transport characteristic of a generic stellarator. In terms of the temperature, this leads to a neoclassical heat flux with a much less unfavorable scaling, proportional to $T^{1/2}$ instead of $T^{9/2}$ \cite{beidler2011ICNTS}.

The constraints that omnigenity imposes on the variation of $B$ on the flux surface are best understood if the latter is parametrized using Boozer poloidal and toroidal angles, $\theta$ and $\zeta$~\cite{boozer1983qs}. In these coordinates, $\mathrm{d}l/  \mathrm{d\theta} = (B_\zeta/\iota+B_\theta)/B$, where $B_\theta = \mathbf{B}\cdot \mathbf{e_\theta}$, $B_\zeta = \mathbf{B}\cdot \mathbf{e_\zeta}$, and the rotational transform $\iota$ are flux-surface constants. It then becomes clear that the spatial dependence of the integrand of {$J$} is determined by $B(\theta,\zeta)$. In an axisymmetric tokamak, $B=B(\theta)$ guarantees that $J$ is a flux-surface constant. Conversely, in quasisymmetric (QS) stellarators, a subset of omnigenous fields, this is achieved by making $B$ depend on the angular position through $M\theta-N\zeta$, $N$ and $M$ being two integer numbers~\cite{boozer1983qs,nuhrenberg1988qs}. Notably, in the $N=0$ case, termed `quasi-axisymmetric' (QA), $B=B(\theta)$ is achieved in the absence of axisymmetry. Finally, more general omnigenous fields exist \cite{cary1997omni,parra2015omni}, in which $M\theta-N\zeta$ is not a direction of symmetry, but in which constancy of $J$ on the flux surface imposes qualitatively similar constraints: in particular, all the contours of {constant} $B$ on the flux surface must close in the $M\theta-N\zeta$ direction. Omnigenous fields, QS or not, are thus said to display a particular \textit{helicity}.

The helicity plays a key role in the size of the bootstrap current, which, if large enough, can modify the magnetic configuration~\cite{peeters2000bootstrap}, with consequences for heat and particle exhaust. The island divertor, arguably the most mature stellarator exhaust concept~\cite{sunnpedersen2019divertor}, exploits resonant magnetic islands at the plasma edge, which requires a vanishing toroidal current. The bootstrap current is zero in omnigenous fields with poloidally-closed $B$-contours (termed quasi-isodynamic, QI)~\cite{helander2009bootstrap}, but is large otherwise~\cite{landreman2012omni}.  For this reason, among the magnetic configurations proposed in recent years as reactor candidates~\cite{sanchez2023qi,goodman2024squids,hegna2025infinity2,lion2025stellaris,swanson2026helios,sanchez2026qi}, QA fields have been much less favored than QI fields, despite the natural advantages of the former.

After all, among omnigenous stellarators, the configuration that lies closest to the tokamak is the QA configuration. The $B=B(\theta)$ dependence that in a tokamak is the straightforward result of axisymmetry (and can be produced with planar coils, arranged with toroidal symmetry, complemented with the inductive current) can be obtained in QA stellarators with relatively small shaping of the flux surfaces. Indeed, a QA stellarator can be viewed analytically as a tokamak subject to a controlled three-dimensional perturbation, see e.g.~\cite{plunk2020qas}. Numerically, acceptable levels of quasisymmetry have been obtained~\cite{moroz1998dhs} replacing the centre post of a standard spherical tokamak coil system with a helical winding. More recently, similar configurations with higher level of quasisymmetry have been produced with tokamak-like coils complemented with banana coils in the inboard side~\cite{henneberg2024hybrid} or with small planar saddle coils placed on a surface surrounding the plasma~\cite{gates2025planar}. Finally, quasi-axisymmetry also brings additional advantages, such as the possibility of screening of impurities \cite{connor1973screening} and of large plasma flows \cite{sugama2011flows} that may suppress turbulent transport.

If a nearly QA stellarator compatible with an island divertor could be realized, it would combine the best features of the three concepts (tokamak, QA and QI), offering good confinement and relative coil simplicity, together with operation robustness and a proven divertor solution. Until now, this combination was regarded as fundamentally impossible due to the unavoidable large bootstrap current associated with toroidal helicity. In this Letter, we propose a strategy to overcome this long-standing obstruction{: we demonstrate that adding a piecewise omnigenous perturbation to a QA field makes it possible to control the bootstrap current level without adversely affecting radial neoclassical transport.}

\section{Theoretical framework}\label{SEC_THEORY}

Piecewise omnigenous (pwO) fields~\cite{velasco2024pwO,velasco2025parapwO,calvo2025pwO,velasco2026pwO} have been recently introduced as an alternative to omnigenity for stellarator optimization~\cite{fernandezpacheco2026pw1,liu2025omni,liu2026opwO}. In pwO fields, tokamak-like neoclassical {transport ~\cite{velasco2024pwO} is} achieved thanks to $J$ being {piecewise} constant, rather than constant, on the flux surface (see \cite{velasco2024pwO,calvo2025pwO,calvo2026pwO}).  Through equation (\ref{EQ_J}), this imposes different requirements to the variation of $B$ on the flux surface. In particular, the $B$-contours of pwO fields do not need to close in the toroidal, helical or poloidal direction. Therefore, pwO fields lack a defined helicity, and could actually be thought to combine several helicities on {the} same flux surface, as we will see later. This has been shown in \cite{calvo2025pwO} to be key for controlling the size of the bootstrap current in \textit{prototypical} pwO fields (and, in particular, its vanishing). 

In this section, we will add a pwO perturbation to an exactly QA field. Specifically, we will design fields, termed QA-pwO, in which deeply trapped particles will behave as in a QA field, with bounce points lying on horizontal $B$-contours, and barely trapped particles moving as in a pwO field, with bounce points that lie on the sides of a parallelogram. Building on the theoretical work of~\cite{velasco2024pwO,velasco2025parapwO,calvo2025pwO} we will choose the shape of the pwO region that minimizes the bootstrap current while keeping a reduced level of radial neoclassical transport.

\begin{figure}
\includegraphics[angle=0,width=1.0\columnwidth]{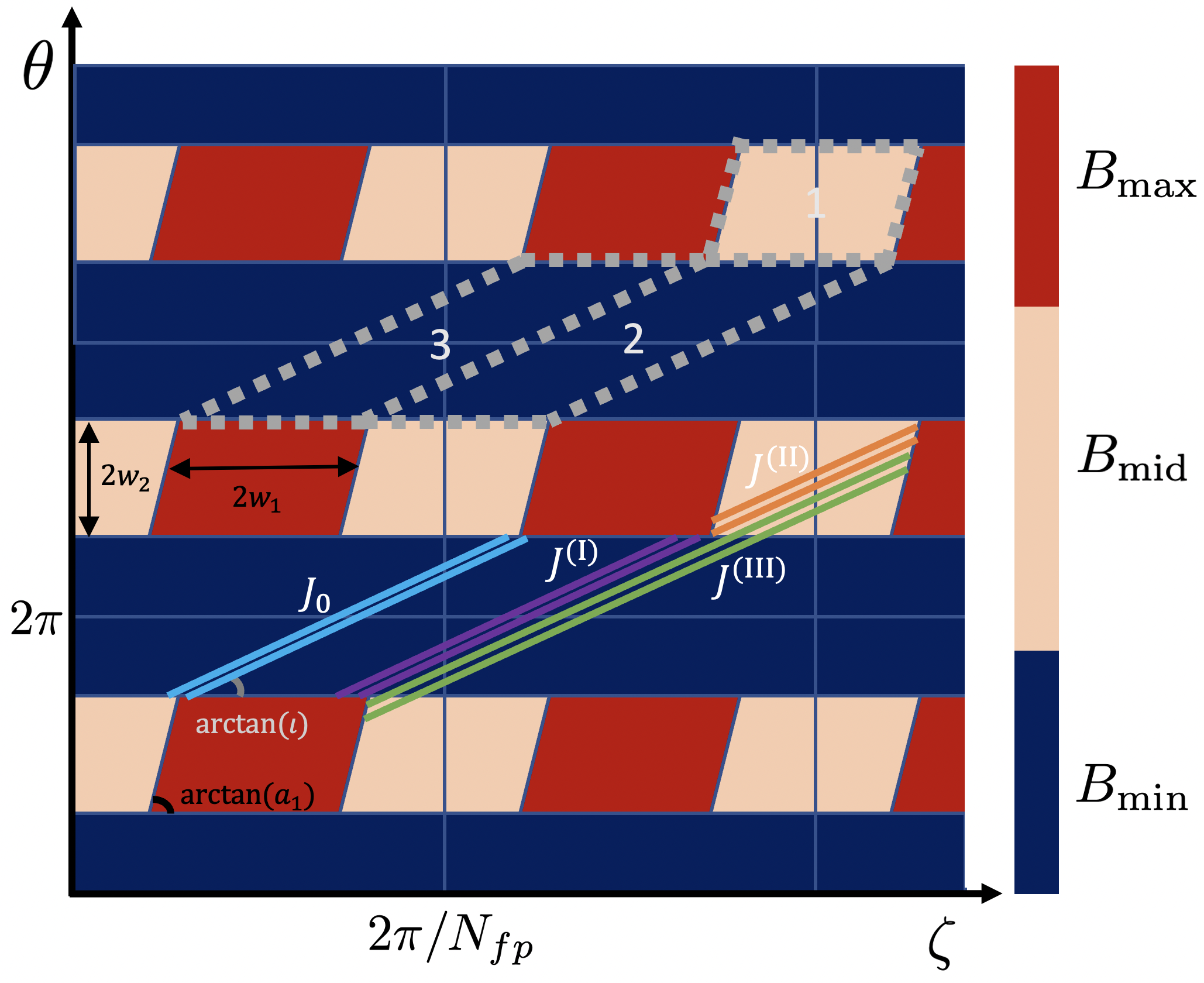}
\caption{$B$ on the flux surface of a prototypical QA-pwO field. For $B_\mathrm{mid}=B_\mathrm{max}$, the field becomes QA.\label{FIG_ANA}}
\end{figure}

Figure~\ref{FIG_ANA} shows an example of a prototypical QA-pwO, field in which the bootstrap current can be computed analytically (a more realistic example will be presented in section~\ref{SEC_RESULTS}). The magnetic field strength displays three different values. Within a period, for $\theta{<}\pi-w_2$ or $\theta{>}\pi+w_2$, $B=B_\mathrm{min}$. For $\pi-w_2\le\theta\le\pi+w_2$, the flux surface is divided into two parallelograms (one of which lies on two different field periods). Their sides have, in both cases, slope 0 and $a_1$. The toroidal extent of the parallelogram with $B=B_\mathrm{max}$ is $2w_1$. The toroidal extent of the one with $B=B_\mathrm{mid}$ is $2(\pi/N_{fp}-w_1)$, $N_{fp}$ being the number of field periods. {The rotational transform $\iota=(i_\theta\pi-w_2)/(i_\zeta\pi/N_{fp}-w_2/a_1)$} ensures that the top corners of any $B=B_\mathrm{max}$ parallelogram are connected by only two field lines with the bottom corners of another $B=B_\mathrm{max}$ parallelogram, with $i_\zeta$ and $i_\theta$ integer numbers and $i_\zeta 2\pi/N_{fp}$ and $i_\theta 2\pi$ the toroidal and poloidal distance between the centers of the connected {parallelograms~\cite{velasco2025parapwO}. The} fact that the $B=B_\mathrm{max}$ parallelogram is centered at $(\zeta,\theta)=(\pi/N_{fp},\pi)$ ensures stellarator symmetry. Crucially, {the expression for $\iota$} does not contain $w_1$. 
 
From the point of view of radial transport, the phase space can be divided into four relevant types of orbits. Trapped particles with $B_\mathrm{min}<\mathcal{E}/\mu \le B_\mathrm{mid}$ (light blue) have their bounce points at $\theta=\pi\pm w_2$ and behave as in a QA field, with $J=J_0$. Conversely, trapped particles with $B_\mathrm{mid}<\mathcal{E}/\mu \le B_\mathrm{max}$ see a pwO field, and may have three different values of $J$: $J^\mathrm{(I)}$ (purple), $J^\mathrm{(II)}$ (orange) and $J^\mathrm{(III)}=J^\mathrm{(I)}+J^\mathrm{(II)}$ (green). This spatial distribution of $J$ ensures tokamak-like radial neoclassical transport of the bulk species at low collisionality.

From the point of view of parallel transport, there are three relevant regions. The $B=B_\mathrm{mid}$ region is labelled as region 1. Region 2 is a parallelogram with two sides of slope $\iota$ and two horizontal sides lying along the boundaries between $B = B_\mathrm{min}$ and $B = B_\mathrm{mid}$. The remaining portion of the $B = B_\mathrm{min}$ area is labeled as region 3.


As explained in Appendix~\ref{SEC_CALC}, the bootstrap current of the magnetic field of figure~\ref{FIG_ANA} is zero if 
\begin{eqnarray}
\Delta_\mathrm{QApwO}=0\,,
\label{EQ_DELTA0}
\end{eqnarray}
where\vspace{-0.2cm}
\begin{eqnarray}
\Delta_\mathrm{QApwO}=\frac{A_{1}}{(\iota-a_1)B_\mathrm{mid}^2} \left[\hat f_t(B_\mathrm{mid},B_\mathrm{max})-f_t\right]+\nonumber  \\
\frac{A_{2}}{(\iota-a_1)B_\mathrm{min}^2}\left[\hat f_t(B_\mathrm{min},B_\mathrm{max})-\hat f_t(B_\mathrm{min},B_\mathrm{mid})\right]+ \nonumber \\
\hspace{-0.5cm}\frac{A_{2}}{\iota B_\mathrm{min}^2}\left[\hat f_t(B_\mathrm{min},B_\mathrm{mid})-f_t\right]+
\frac{A_{3}}{\iota B_\mathrm{min}^2}\left[\hat f_t(B_\mathrm{min},B_\mathrm{max})-f_t\right].
\label{EQ_DELTA}
\end{eqnarray}
The quantities $\hat f_t$ and $f_t$ take $O(1)$ values, and $A_i$ is the area in ($\theta,\zeta$) space of region $i$. The definitions of $\hat f_t$ and $f_t$, and expressions for $A_i$ are provided in Appendix~\ref{SEC_DEF}.

Equation (\ref{EQ_DELTA0}) needs to be solved numerically. One of the solutions is $N_{fp}=2$, $i_\zeta=4$, $i_\theta=1$, $\iota=0.439$, $a_1=0.616$, $w_1=0.559\pi/N_{fp}$, $w_2=0.424\pi$, $B_\mathrm{min}=0.9\,$T, $B_\mathrm{mid}=1.024\,$T, {$B_\mathrm{max}=1.1\,$T}. This solution, along with the discussion of this section, serves the purpose of illustrating key aspects required to achieve a reduced bootstrap current. $B$-contours of constant $\theta$ give a contribution to the bootstrap current with a definite sign (in a QA field, only this type of $B$-contours exist, giving a large bootstrap current comparable to that of the tokamak). In order to compensate for this, there must exist $B$-contours of constant $\theta-a_1\zeta$, with negative $\iota-a_1$. The relative areas of regions 1, 2 and 3, combined with the relative values of $B$, of $\iota$ and of $a_1-\iota$ will determine whether the bootstrap current vanishes at low collisionality. 

{We end this section by noting that this framework can be smoothly extended to quasi-helically symmetric fields.}

\section{A smooth nearly QA field with reduced bootstrap current}\label{SEC_RESULTS}

\begin{figure}
\includegraphics[angle=0,width=0.49\columnwidth]{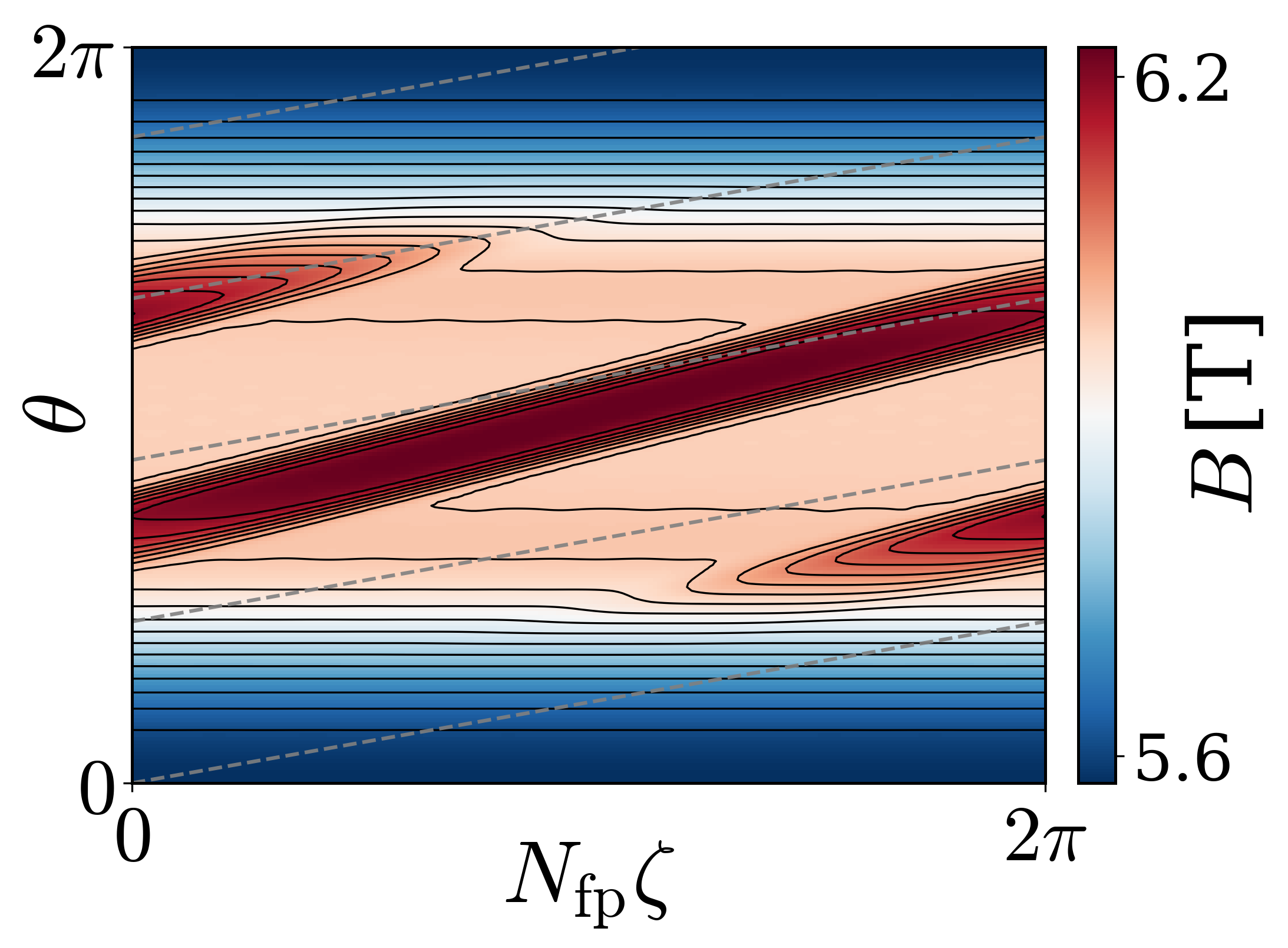}
\includegraphics[angle=0,width=0.49\columnwidth]{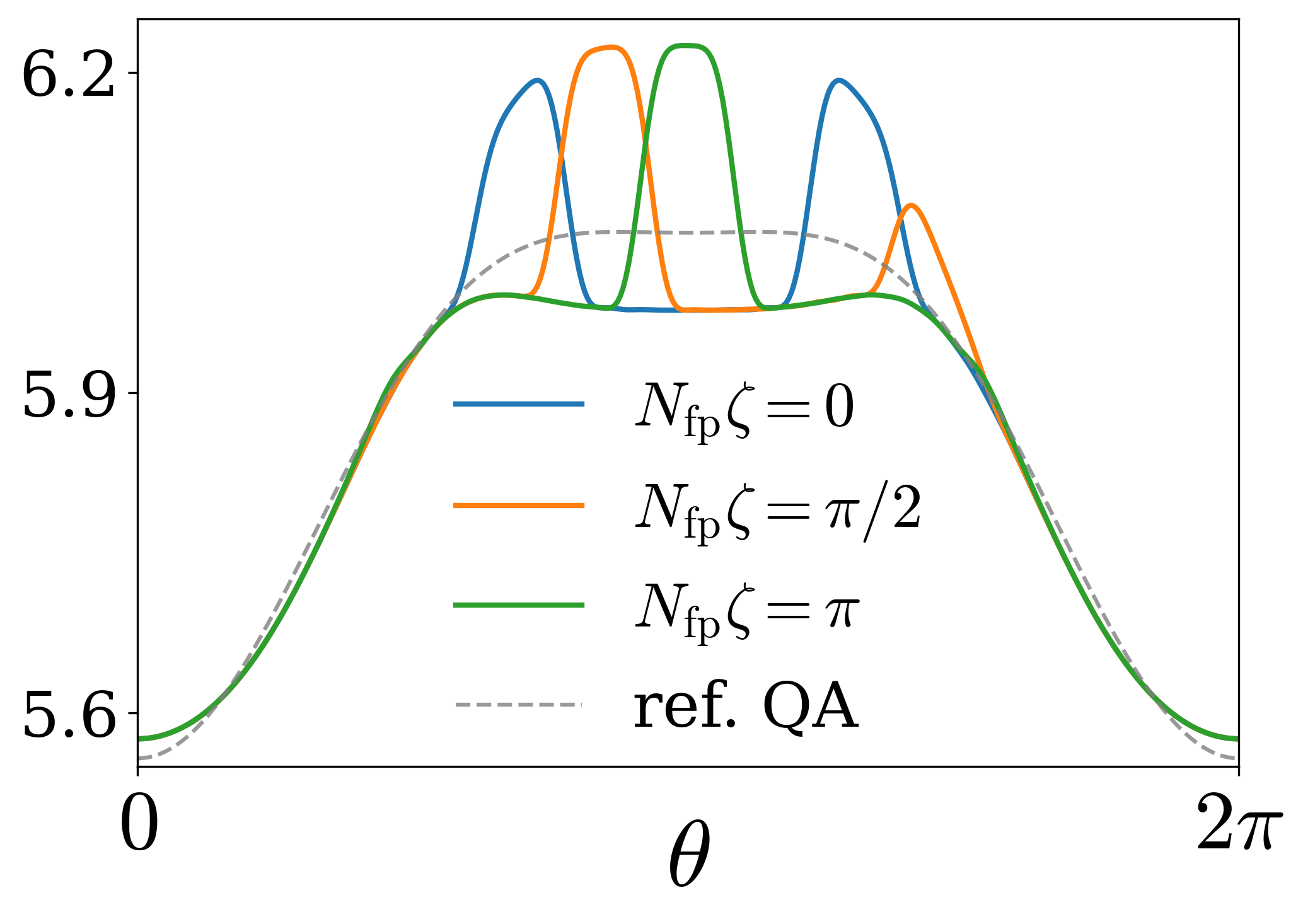}
\caption{$B$ on the flux surface of a smooth nearly QA field with a pwO perturbation that reduces the bootstrap current.\label{FIG_MOD}}
\end{figure}

Figure~\ref{FIG_MOD} shows the magnetic field strength of a model QA-pwO field, similar to that of figure~\ref{FIG_ANA} (so that its features can still be discussed in terms of quantities such as $w_1$ or $a_1$), but more realistic. {Figure~\ref{FIG_MOD} (right) also depicts an exactly QA field that can be used for direct comparison: differences of the order of 3\% can be observed close to the maxima. The parameters describing $B(\theta,\zeta)$ and the reference QA field, including the Fourier spectrum, are provided in Appendix~\ref{SEC_FOURIER}. }

\begin{figure}
\includegraphics[angle=0,width=\columnwidth]{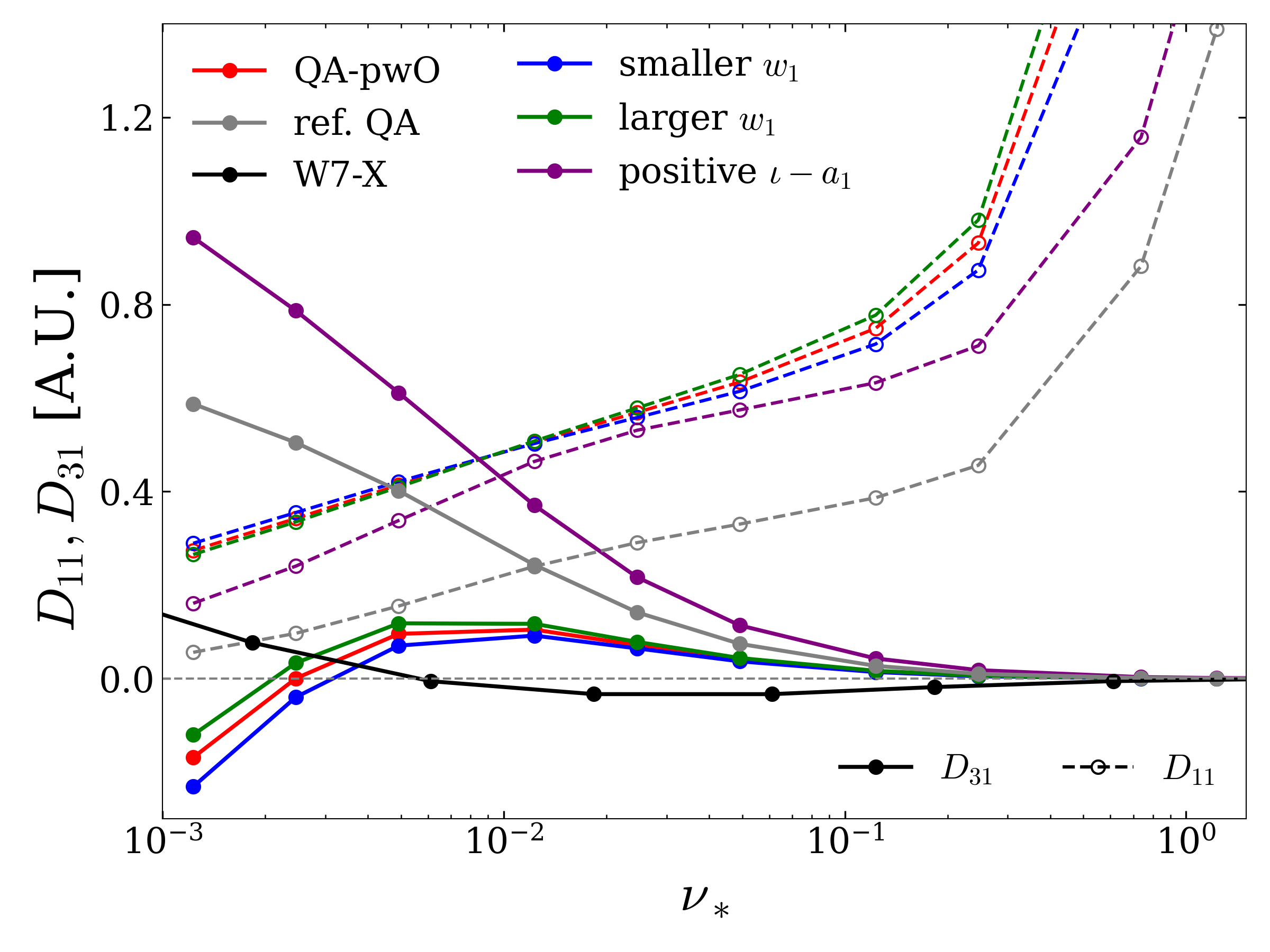}
\caption{Radial ($D_{11}$) and parallel ($D_{31}$) neoclassical transport coefficients as a function of the collisionality.\label{FIG_NC}}
\end{figure}

Figure~\ref{FIG_NC} shows the level of parallel and radial transport of several QA-pwO magnetic fields. Specifically, the $D_{11}$ and $D_{31}$ neoclassical transport coefficients are computed with \texttt{MONKES}~\cite{escoto2024monkes,escoto2025QI}, as these coefficients encapsulate the dependence of neoclassical transport on the details of the magnetic configuration, see e.g.~\cite{beidler2011ICNTS} {(only the calculations in the absence of a radial electric field $E_r$ are shown, since very weak dependence is observed with $E_r$)}. The main thesis of this work is proven correct: a pwO perturbation can significantly reduce the bootstrap current (with respect to a reference QA, compare red and gray) while keeping a low radial neoclassical transport.  Specifically, at reactor-relevant collisionality, the bootstrap current can be made to vanish without causing the emergence of a $1/\nu$ regime for radial transport {(an effective ripple~\cite{beidler2011ICNTS} around $10^{-4}$ can be estimated computing $D_{11}$ at even lower collisionalities)}. Additionally, the QA-pwO cases with a smaller/larger value of $w_1$ (blue/green) illustrate some degree of control over the bootstrap current level.

{Interestingly, t}he case with positive $\iota-a_1$ (purple) shows that the bootstrap current can be increased, rather than reduced. In tokamaks, carefully designed QS deviations from axisymmetry have been experimentally shown to constitute a robust path for error field correction~\cite{park2021qs}. Similarly, pwO deviations from axisymmetry, if they could be produced without significant toroidal flow damping, could represent a viable strategy towards tokamak operation with a higher bootstrap current fraction.

\section{Discussion}\label{SEC_DISCUSSION}

The result of section~\ref{SEC_RESULTS} opens a new pathway for the design of a stellarator reactor that {potentially combines simple coils, tokamak-like confinement, and compatibility with an island divertor. However, substantial work remains to fully explore its potential}. In this section we present a few preliminary analyses in this direction.

In figure~\ref{FIG_NC}, the QA-pwO field of figure~\ref{FIG_MOD} showed, at low collisionalities, a bootstrap level roughly an order of magnitude below a QA field. This is a reduction comparable {to that} of the \textit{standard} configuration of W7-X {(appropriately scaled in figure~\ref{FIG_NC})}, which has been experimentally demonstrated to be compatible with an island divertor~\cite{sunnpedersen2019divertor}. However, when scaled to a reactor scenario with higher temperatures, this level of bootstrap current may still be too high. Further optimization, in order to make $D_{31}$ comparable to that of the high-mirror configuration of W7-X, could then be necessary~\cite{beidler2021nature,beidler2001hsr4}. In order to assess this, we next give an order-of-magnitude estimate of the effect of \jl{the bootstrap level} of figure~\ref{FIG_NC}.

\begin{figure}
\includegraphics[angle=0,width=\columnwidth]{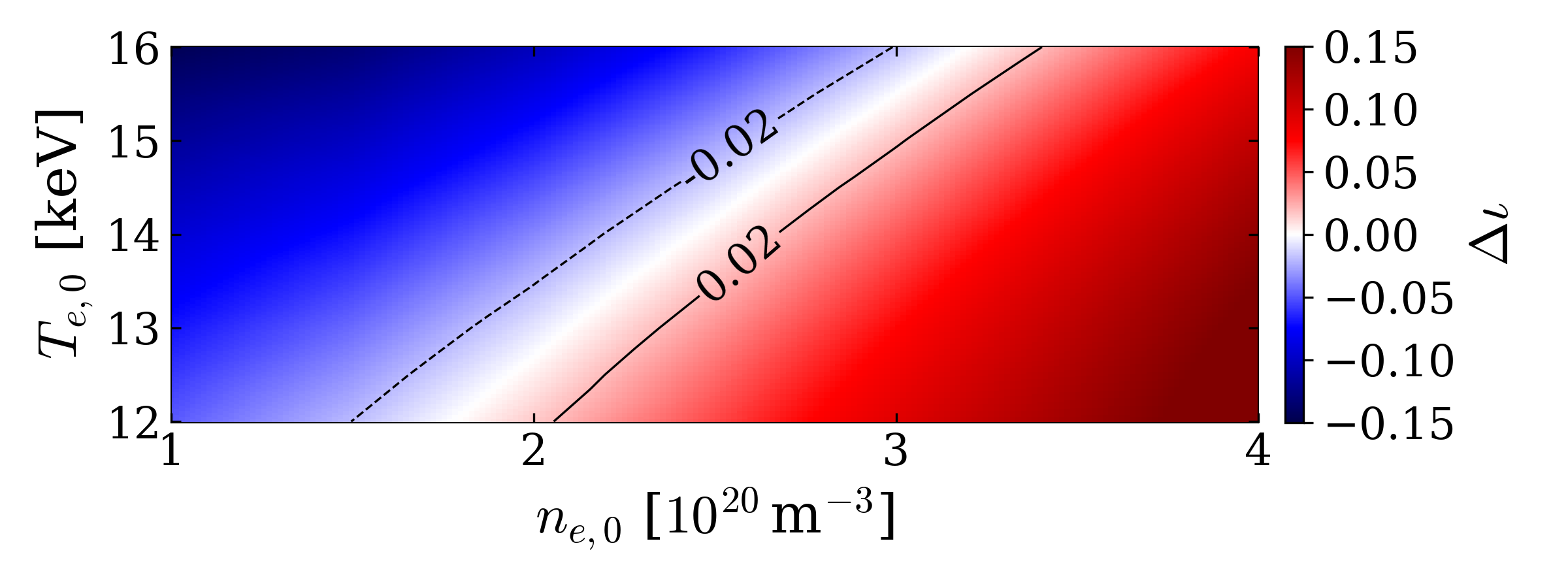}
\caption{Change in the rotational transform due to the bootstrap current in a reactor scenario for a model QA-pwO field.\label{FIG_I}}
\end{figure}

We compute the bootstrap current profile of a reactor scenario, and use it to estimate the associated change in the rotational transform at the edge, $\Delta\iota$. Details on the calculation, including the plasma profiles and how the model fields (which are defined on a flux surface) are extrapolated to a whole volume are given in Appendix~\ref{SEC_REACTOR}. The results are shown in figure \ref{FIG_I} as a function of the density and temperature at the core. Even if $D_{31}$ is too large, the fact that it changes sign at low collisionality already offers a pathway for the reduction $\Delta\iota$ to acceptable levels. In particular, $\Delta\iota=0$, required for an island divertor, can result from canceling a negative core contribution (where $\nu_*$ is lower) with a positive edge contribution.

The range of plasmas for which $\Delta\iota$ is small enough (the contours $\Delta\iota=\pm 0.02$ are highlighted in figure~\ref{FIG_I}, as a rough indicator of the variation of $\iota$ that can be expected from other finite-$\beta$ effects) is determined by the size of $D_{31}$ at intermediate collisionalities. For the model QA-pwO field, this range is comparable with the one that could be obtained for a reactor based on the standard configuration of W7-X. On the one hand, this means that further reduction of $D_{31}$ at intermediate collisionalities would be desirable for the safe operation of an island divertor. On the other hand, the finding of this Letter represents a fundamental advancement, as no reactor plasmas exist that yield $\Delta\iota\approx 0$ in a QA stellarator.

The need for scenario development is not new in stellarators (QI candidates~\cite{regana2025qi,goodman2024squids,hegna2025infinity2,lion2025stellaris,sanchez2026qi}, for instance, rely for turbulence reduction on a relatively fine~\cite{bozhenkov2026routes} control {over} the density profile shape). Since the temporal evolution of the toroidal current is slower than the transport time scale~\cite{nrl2018LRtime}, a relatively small (compared to a QA field) $\Delta\iota$ could perhaps be compensated by adjustments of the average density, or by current drive. In any case, further optimization of the (self-consistent \cite{feng2020bootstrap,landreman2022bootstrap}) bootstrap current should undoubtedly be the next step.

\begin{figure}
\includegraphics[angle=0,width=\columnwidth]{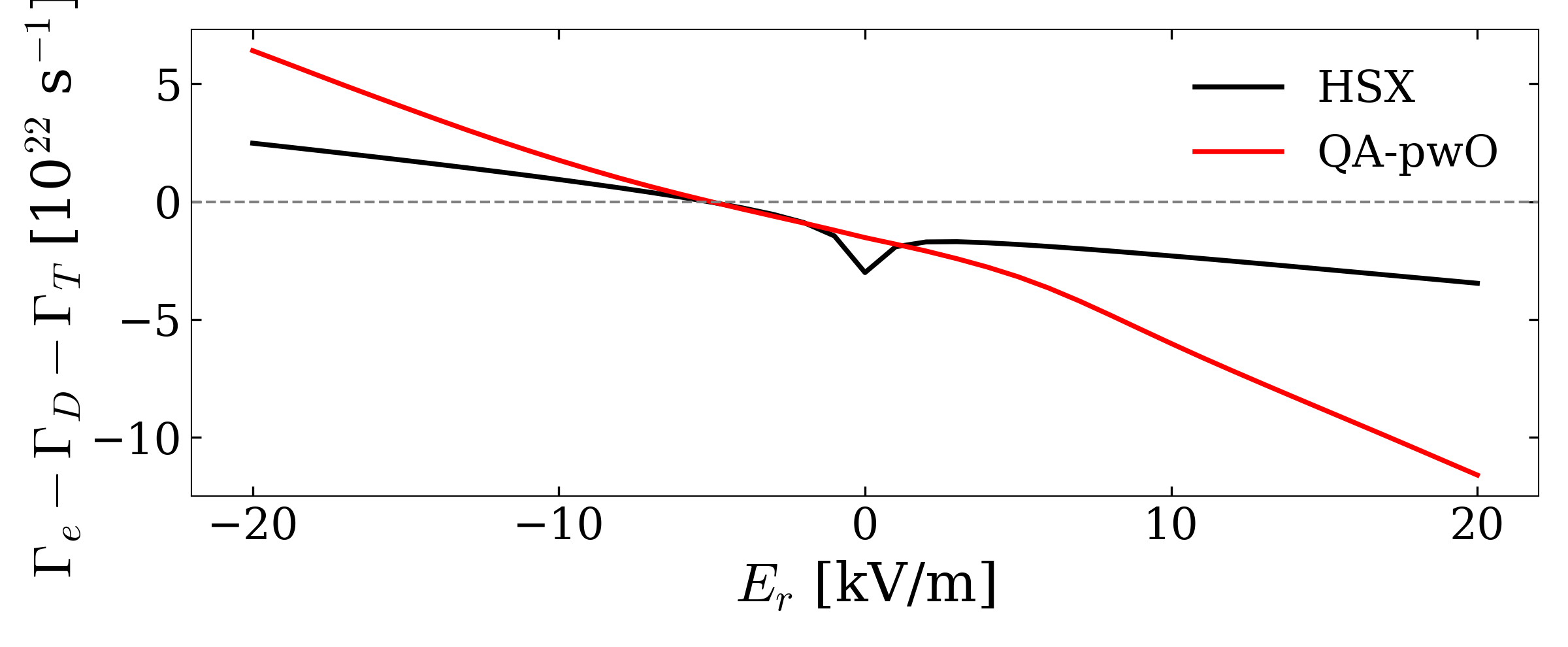}
\caption{Dependence of the radial charge current on the radial electric field $E_r$ for $n_{e,0}=4\times 10^{20}\,$m$^{-3}$ and $T_{e,0}=12\,$keV.\label{FIG_AMB}}
\end{figure}

The fields presented in this Letter can be argued to be approximately quasisymmetric. However, particles with $\mathcal{E}/\mu\lesssim B_\mathrm{max}$ do experience {relatively large} deviations from quasisymmetry, which casts doubts on the capability of such fields to sustain large flows (or equivalently, to be automatically ambipolar \cite{calvo2014er,calvo2015flowdamping}) or screen impurities. Since earlier analytical studies \cite{calvo2013er,calvo2014er,calvo2015flowdamping} are not directly applicable, we address the problem by means of a numerical comparison with the approximately quasisymmetric stellarator HSX scaled to reactor size, whose level of optimization has been experimentally proven to reduce radial transport and flow damping in the direction of symmetry \cite{gerhardt2005qhs,canik2007qhs} (the behaviour of impurities, on the other hand, is qualitatively similar to that of a non-QS stellarator~\cite{swee2022hsx}). Figure~\ref{FIG_AMB} shows that the model QA-pwO field and HSX (at {$s=0.15$}) \jl{fail to satisfy} the condition of automatic ambipolarity ($\Gamma_e-\Gamma_H-\Gamma_D=0$ for any radial electric field, $\Gamma_b$ being the neoclassical particle flux) to a similar extent (i.e. with a similar {non-zero} slope around {a single} ambipolar $E_r$ value, {$2.8\times 10^{22}$ vs $1.7\times 10^{22}$ m$\,$s$^{-1}$kV$^{–1}$ respectively}). This property may be optimized in QA-pwO {fields} by making the pwO region smaller, in combination with a smaller value of $|\iota-a_1|$.

\begin{figure}
\includegraphics[angle=0,width=0.49\columnwidth]{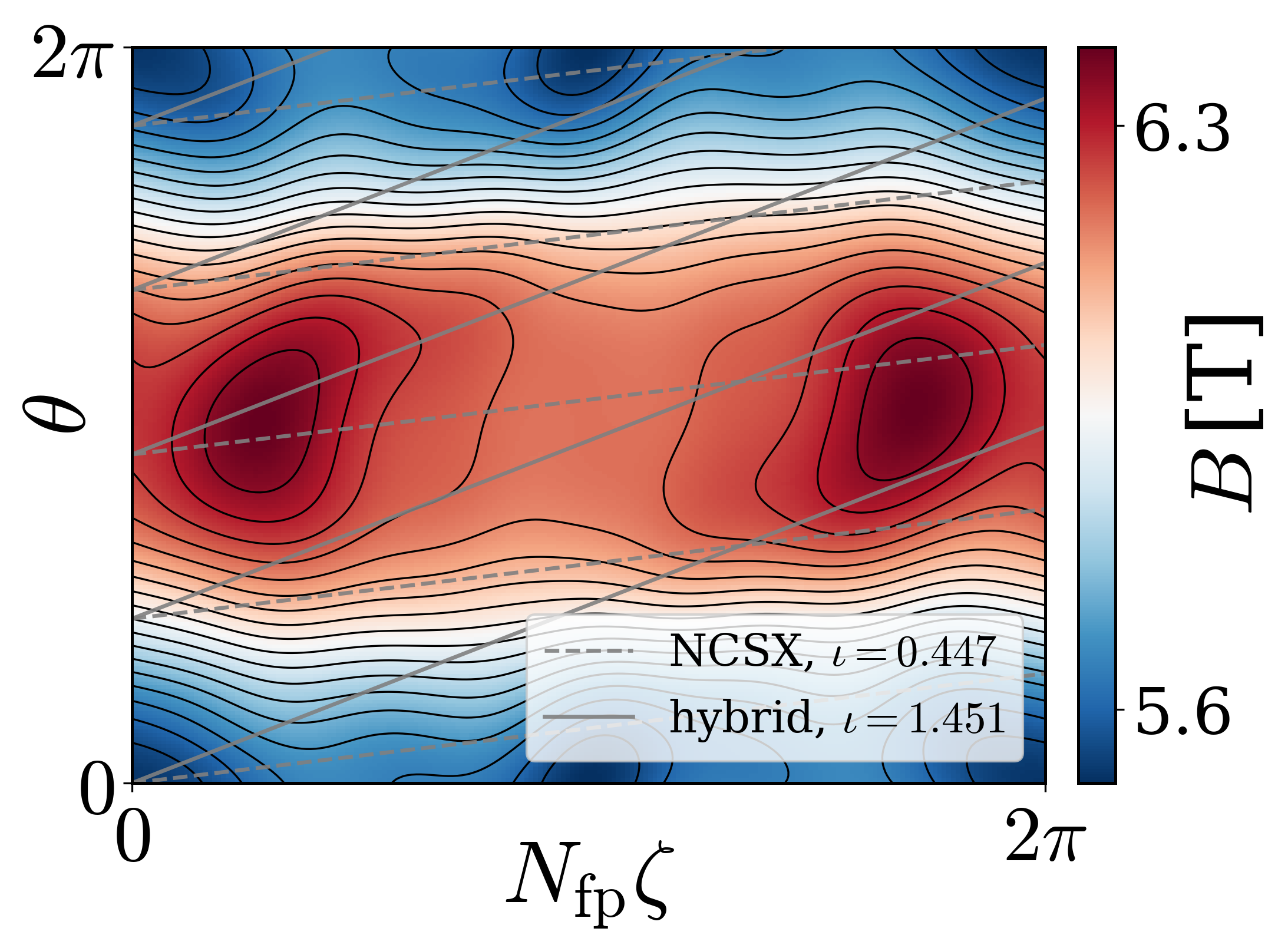}
\includegraphics[angle=0,width=0.49\columnwidth]{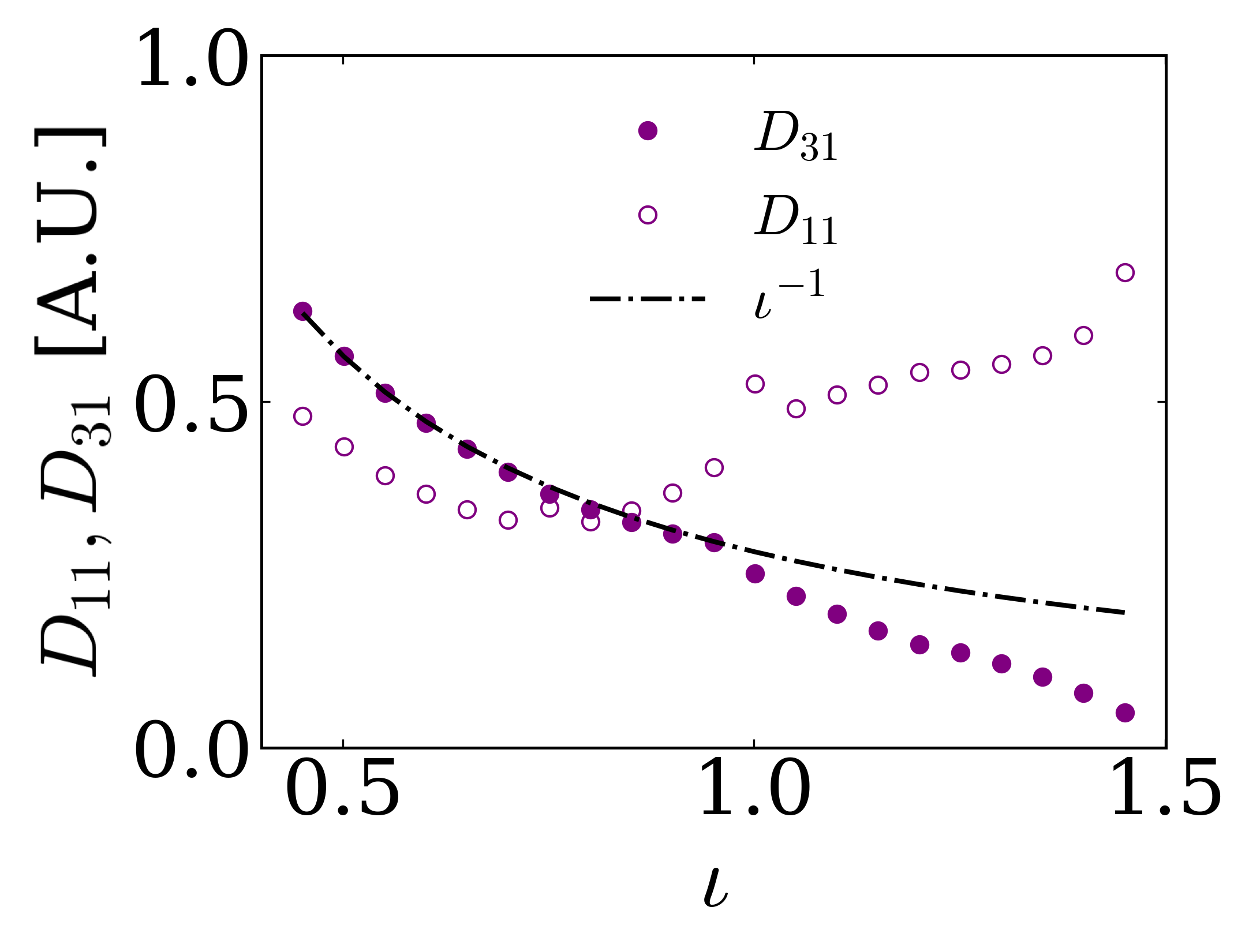}
\caption{{Magnetic-field strength and field lines of NCSX and a tokamak-stellarator hybrid with the same $B$ and different $\iota$ (left), and neoclassical transport \jl{coefficients} for $\nu_*= 10^{-4}$ in a configuration scan between these two cases (right). \label{FIG_NCSX}}}
\end{figure}

{Finally, we discuss the feasibility in fusion-relevant magnetohydrodynamic equilibria of the concept presented in this Letter by constructing a series of tokamak-stellarator hybrids~\cite{hartwell2017hybrid,li2026hybrid,henneberg2024hybrid,halpern2026hybrid} with varying $\iota$. We do so by using the National Compact Stellarator eXperiment (NCSX) as a starting point. The magnetic configuration of NCSX~\cite{zarnstorff2001ncsx} displays, at the same time, relatively large (approximately parallelogram-shaped) deviations from quasisymmetry around $B_\mathrm{max}$, banana radial transport for a range of collisionalities, and an `intriguing'~\cite{beidler2011ICNTS} behaviour of $D_{31}$. It is also the basis of the ARIES-CS reactor design~\cite{ku2008ariescs} and, thanks to the dramatic improvement in the numerical tools for coil design, it can be e.g. created with a simple magnet setup~\cite{zhu2020pm} composed of planar coils and permanent magnets (which could be replaced by planar saddle coils~\cite{gates2025planar}). Figure~\ref{FIG_NCSX} (left) presents the magnetic-field strength and several field lines at $s=0.2$ of NCSX (scaled to reactor size) and one of the hybrids. Both have the same $B(\theta,\zeta)$ and, as a result of an additional toroidal current driven through the plasma, the hybrid has larger $\iota$. Figure~\ref{FIG_NCSX} (right) shows a scan between these two configurations, the effect of which is to bring $\iota$ closer to the value of $a_1$ that can be inferred from the $B$-contours around $B_\mathrm{max}$. As a result of this, in agreement with section~\ref{SEC_THEORY}, the $D_{31}$ coefficient drops significantly below the $\iota^{-1}$ trend of a QA field without $D_{11}$ significantly increasing.}

{In summary, the theoretical framework presented in this Letter, combined with the existence of more efficient optimization tools, can be expected to open a new promising route towards the stellarator reactor.}

\begin{acknowledgments}


This work has been carried out within the framework of the EUROfusion Consortium, funded by the European Union via the Euratom Research and Training Programme (Grant Agreement No 101052200 EUROfusion). Views and opinions expressed are however those of the author(s) only and do not necessarily reflect those of the European Union or the European Commission. Neither the European Union nor the European Commission can be held responsible for them. This research was supported by grants PID2021-123175NB-I00 and PID2024-155558OB-I00, Ministerio de Ciencia, Innovaci\'on y Universidades, Spain. The authors are grateful to all the developers of the codes \texttt{MONKES}, \texttt{Neotransp} and \texttt{SFINCS}. Calculations for this work made use of computational resources at Xula (CIEMAT) and Raven (MPG). 

\end{acknowledgments}

\bibliography{QIpwO.bbl}

\appendix

\

\

\

\large \textbf{Supplemental material} \normalsize

\

\section{Calculation of the bootstrap current for a prototypical QA-pwO field}\label{SEC_CALC}

In order to streamline the calculation, in this appendix we follow the approach and notation of~\cite{helander2017jpp} for the computation of the bootstrap current in large aspect ratio stellarators at low collisionality, using a model collision operator. Then, we point out how, in the spirit of \cite{calvo2025pwO}, one can derive the condition for zero bootstrap current for QA-pwO fields, and therefore obtain equation (\ref{EQ_DELTA0}).

We write the bootstrap current as
\begin{eqnarray}
\fsa{\mathrm{j}_\parallel\cdot\mathrm{B}} = (f_s + \fsa{u B^2}) K,\label{EQ_SC}
\end{eqnarray}
where $K$ depends on the plasma profile gradients and relatively trivial magnetic geometry properties. Here, $u$ is calculated by solving the differential equation
\begin{eqnarray}
\mathrm{B}\cdot\nabla u = - (\mathrm{B}\times\nabla s)\cdot\nabla B^{-2},
\end{eqnarray}
with $s$ the radial coordinate. If $u$ is set to be zero in the region with $B=B_\mathrm{max}$, it is straightforward to show that
\begin{eqnarray}
u_1&=&\frac{B_\mathrm{max}^{-2}-B_\mathrm{mid}^{-2}}{\Psi_t'} \frac{a_1B_\theta+B_\zeta}{\iota-a_1}\,,\nonumber\\
u_2-u_1&=&\frac{B_\mathrm{mid}^{-2}-B_\mathrm{min}^{-2}}{\Psi_t'} \frac{B_\zeta}{\iota}\,,\nonumber\\
u_3&=&\frac{B_\mathrm{max}^{-2}-B_\mathrm{min}^{-2}}{\Psi_t'} \frac{B_\zeta}{\iota}\,,\label{EQ_U}
\end{eqnarray}
where $2\pi\Psi_T$ is the toroidal flux through the last-closed flux surface and prime denotes radial derivative. $f_s$ is defined as
\begin{eqnarray}
f_s = \frac{3\fsa{B^2}}{4}\int_0^{B^{-1}_\mathrm{max}}\mathrm{d}\lambda \frac{\lambda\fsa{g_4}}{\fsa{\sqrt{1-\lambda B}}}\,,
\end{eqnarray}
with $g_4$ obtained by solving
\begin{eqnarray}
\mathrm{B}\cdot\nabla\left(\frac{g_4}{\xi}\right) = - (\mathrm{B}\times\nabla s)\cdot\nabla \xi^{-1},
\end{eqnarray}
with $\xi=\sigma\sqrt{1-\lambda B}$ and $\sigma=\pm 1$ the sign of the parallel velocity. If $g_4$ is set to be zero in the region with $B=B_\mathrm{max}$, it is straightforward to show that the values of $g_4$ in regions 1, 2 and 3 are given by
\begin{eqnarray}
(g_4)_1&=&\left(1-\frac{\xi_\mathrm{mid}}{\xi_\mathrm{max}}\right)\frac{a_1B_\theta+B_\zeta}{\iota-a_1}\,,\nonumber\\
(g_4)_2-(g_4)_1\frac{\xi_\mathrm{min}}{\xi_\mathrm{mid}}&=&\left(1-\frac{\xi_\mathrm{min}}{\xi_\mathrm{mid}}\right)\frac{B_\zeta}{\iota}\,,\nonumber\\
(g_4)_3&=&\left(1-\frac{\xi_\mathrm{min}}{\xi_\mathrm{max}}\right)\frac{B_\zeta}{\iota}\,.\label{EQ_G}
\end{eqnarray}
It can be shown~\cite{helander2009bootstrap,calvo2025pwO} that the bootstrap current is zero for any plasma gradients if all the terms in equation (\ref{EQ_SC}) that are proportional to $B_\zeta$ vanish. Computing (\ref{EQ_SC}) using equations (\ref{EQ_U}) and (\ref{EQ_G}), this happens when equation (\ref{EQ_DELTA0}) is fulfilled.

We finally note that, by setting $B_\mathrm{mid}=B_\mathrm{min}$, one can recover the result of~\cite{calvo2025pwO} (for the specific case in which two sides of the parallelogram are {horizontal}).

\section{Definitions and explicit expressions for quantities in equation~(\ref{EQ_DELTA})}\label{SEC_DEF}

In section \ref{SEC_THEORY}, the quantities $\hat f_t$ and $f_t$ have been employed. Their definitions are:
\begin{eqnarray}
& &\hat f_t(B_\mathrm{a},B_\mathrm{b}) = \frac{B_\mathrm{a}^2}{B_\mathrm{b}^2}-  \nonumber\\ & &\frac{3\fsa{B^2}}{4}\int_0^{B^{-1}_\mathrm{max}}\mathrm{d}\lambda \frac{\lambda}{\fsa{\sqrt{1-\lambda B}}}\frac{\sqrt{1-\lambda B_\mathrm{a}}}{\sqrt{1-\lambda B_\mathrm{b}}}\,,
\label{EQ_FC}
\end{eqnarray}
where $\fsa{...}$ denotes flux-surface average, and
\begin{eqnarray}
f_t &=& \hat f_t(B_\mathrm{a},B_\mathrm{a})\,.
\end{eqnarray}
We note that $f_t$ and $f_c=1-f_t$ are quantities usually employed in derivations that involve the solution of differential equations on the flux surface of stellarators, see e.g. \cite{calvo2018nf}.

Additionally, the areas in $(\theta,\zeta)$ space of regions 1, 2, and 3, defined in Section~\ref{SEC_THEORY}, can be expressed in terms of the parameters of the pwO piece of the QA-pwO field:
\begin{eqnarray}
A_1 &=& 4w_2(\pi/N_{fp}-w_1)\,,\nonumber\\
A_2 &=& 4(\pi-w_2)(\pi/N_{fp}-w_1)\,,\nonumber \\
A_3 &=& 4(\pi-w_2)w_1\,.
\end{eqnarray}

\section{pwO parametrization and Fourier spectrum of the magnetic field of figure~\ref{FIG_MOD}}\label{SEC_FOURIER}

The pwO region of the magnetic field strength of figure~\ref{FIG_MOD} can be {approximately described} by:
\begin{eqnarray}
N_{fp}&=&2,\nonumber\\
i_\zeta&=&4,\nonumber\\
i_\theta&=&1,\nonumber\\
w_1&=&0.3\pi/N_{fp},\nonumber\\
w_2&=&0.424\pi,\nonumber\\
a_1&=&0.616,\nonumber\\
\iota&=&0.439.
\end{eqnarray}
This is replaced by $w_1=0.35\pi/N_{fp}$ and $w_1=0.25\pi/N_{fp}$ for the cases with increased and reduced $w_1$, respectively.  Finally, for the case with positive $\iota-a_1$, we set $a_1=0.407$ and $\iota=0.6011$.

The Fourier spectrum of the magnetic field of figure~\ref{FIG_MOD} is
\begin{eqnarray}
B(\theta,\zeta) &=& \sum_{m,n} B_{m,n} \cos(m\theta -nN_{fp}\zeta)\,,
\end{eqnarray}
with the modes {$|B_{mn}/B_{00}|>10^{-4}\,$} shown in table \ref{TAB_BMN}. {The reference exactly-QA magnetic field is composed of all the $B_{mn}\,$ modes of table \ref{TAB_BMN} with $n=0$.}

\

\begin{table}[h!]
    \centering
    \begin{tabular}{c@{\hspace{1.cm}}c@{\hspace{1.cm}}c@{\hspace{1.cm}}c}
    \hline
    $m$ & $n$ & $B_{mn}\,$[T] & $B_{mn}/B_{00}$ \\
    \hline
    0 & 0 & 5.8704 & 1 \\
    1 & 0 & -2.4602 $\times 10^{-1}$ & -4.1909 $\times 10^{-2}$ \\
    2 & 0 & -6.6680 $\times 10^{-2}$ & -1.1359 $\times 10^{-2}$ \\
    3 & 1 & 4.3631 $\times 10^{-2}$ & 7.4323 $\times 10^{-3}$ \\
    4 & 1 & -3.7288 $\times 10^{-2}$ & -6.3519 $\times 10^{-3}$ \\
    2 & 1 & -2.6972 $\times 10^{-2}$ & -4.5946 $\times 10^{-3}$ \\
    6 & 2 & 2.4869 $\times 10^{-2}$ & 4.2363 $\times 10^{-3}$ \\
    7 & 2 & -2.4715 $\times 10^{-2}$ & -4.2101 $\times 10^{-3}$ \\
    5 & 1 & 1.4945 $\times 10^{-2}$ & 2.5458 $\times 10^{-3}$ \\
    3 & 0 & 1.3107 $\times 10^{-2}$ & 2.2327 $\times 10^{-3}$ \\
    5 & 2 & -1.2736 $\times 10^{-2}$ & -2.1696 $\times 10^{-3}$ \\
    8 & 2 & 1.2446 $\times 10^{-2}$ & 2.1201 $\times 10^{-3}$ \\
    10 & 3 & -8.2574 $\times 10^{-3}$ & -1.4066 $\times 10^{-3}$ \\
    9 & 3 & 7.1465 $\times 10^{-3}$ & 1.2174 $\times 10^{-3}$ \\
    4 & 0 & 6.6360 $\times 10^{-3}$ & 1.1304 $\times 10^{-3}$ \\
    16 & 5 & 5.2264 $\times 10^{-3}$ & 8.9029 $\times 10^{-4}$ \\
    1 & 1 & 5.0324 $\times 10^{-3}$ & 8.5724 $\times 10^{-4}$ \\
    11 & 3 & 5.0306 $\times 10^{-3}$ & 8.5694 $\times 10^{-4}$ \\
    17 & 5 & -4.4105 $\times 10^{-3}$ & -7.5131 $\times 10^{-4}$ \\
    0 & 1 & 4.3779 $\times 10^{-3}$ & 7.4576 $\times 10^{-4}$ \\
    7 & 1 & -4.1999 $\times 10^{-3}$ & -7.1543 $\times 10^{-4}$ \\
    15 & 5 & -3.2780 $\times 10^{-3}$ & -5.5840 $\times 10^{-4}$ \\
    8 & 3 & -2.9249 $\times 10^{-3}$ & -4.9824 $\times 10^{-4}$ \\
    19 & 6 & 2.8471 $\times 10^{-3}$ & 4.8499 $\times 10^{-4}$ \\
    20 & 6 & -2.7946 $\times 10^{-3}$ & -4.7605 $\times 10^{-4}$ \\
    3 & 2 & 2.7577 $\times 10^{-3}$ & 4.6977 $\times 10^{-4}$ \\
    10 & 2 & -2.7518 $\times 10^{-3}$ & -4.6876 $\times 10^{-4}$ \\q
    13 & 4 & 2.7180 $\times 10^{-3}$ & 4.6300 $\times 10^{-4}$ \\
    1 & -1 & -2.6122 $\times 10^{-3}$ & -4.4499 $\times 10^{-4}$ \\
    12 & 4 & -2.0135 $\times 10^{-3}$ & -3.4299 $\times 10^{-4}$ \\
    14 & 4 & -1.9607 $\times 10^{-3}$ & -3.3399 $\times 10^{-4}$ \\
    18 & 5 & 1.7303 $\times 10^{-3}$ & 2.9476 $\times 10^{-4}$ \\
    6 & 1 & 1.6914 $\times 10^{-3}$ & 2.8812 $\times 10^{-4}$ \\
    5 & 0 & -1.6352 $\times 10^{-3}$ & -2.7855 $\times 10^{-4}$ \\
    18 & 6 & -1.4826 $\times 10^{-3}$ & -2.5255 $\times 10^{-4}$ \\
    21 & 6 & 1.3834 $\times 10^{-3}$ & 2.3565 $\times 10^{-4}$ \\
    2 & 2 & -1.0310 $\times 10^{-3}$ & -1.7563 $\times 10^{-4}$ \\
    11 & 2 & 9.8789 $\times 10^{-4}$ & 1.6828 $\times 10^{-4}$ \\
    12 & 3 & -8.9007 $\times 10^{-4}$ & -1.5162 $\times 10^{-4}$ \\
    13 & 3 & -8.3903 $\times 10^{-4}$ & -1.4293 $\times 10^{-4}$ \\
    8 & 1 & 8.3863 $\times 10^{-4}$ & 1.4286 $\times 10^{-4}$ \\
    6 & 3 & 8.0607 $\times 10^{-4}$ & 1.3731 $\times 10^{-4}$ \\
    4 & 2 & 7.7581 $\times 10^{-4}$ & 1.3216 $\times 10^{-4}$ \\
    26 & 8 & 7.1962 $\times 10^{-4}$ & 1.2258 $\times 10^{-4}$ \\
    14 & 5 & 6.4289 $\times 10^{-4}$ & 1.0951 $\times 10^{-4}$ \\
    9 & 1 & 6.3812 $\times 10^{-4}$ & 1.0870 $\times 10^{-4}$ \\
    11 & 4 & 6.1547 $\times 10^{-4}$ & 1.0484 $\times 10^{-4}$ \\
    9 & 2 & -6.1488 $\times 10^{-4}$ & -1.0474 $\times 10^{-4}$ \\
    \hline
    \end{tabular}
\caption{Fourier modes of the field of figure~\ref{FIG_MOD}.}\label{TAB_BMN}
\end{table}

{For these fields, the resolution required by \texttt{MONKES} to achieve a $5\%$ precision in the calculation of $D_{31}$ is $N_\theta=70$, $N_\zeta=70$, $N_\xi=200$. This is the result of a convergence scan analogous to that of~\cite{escoto2024monkes}.}

\section{Reactor scenario}\label{SEC_REACTOR}

In this section, we describe the steps required to produce reactor-relevant neoclassical calculations out of model fields $B_\mathrm{QApwO}(\theta,\zeta)$ defined on a flux surface. We will use $0\le s \le 1$, proportional to the toroidal flux, as  the radial coordinate.

First, inspired by~\cite{fernandezpacheco2026pw1}, we extrapolate the model fields radially {using different prescriptions for $s < s_0$ and $s > s_0$:}
\begin{eqnarray}
B(s,\theta,\zeta) &=& \sum_{m,n} B_{m,n}(s) \cos(m\theta -nN_{fp}\zeta)\,,\\
B_{m,n}(s)  &=& B_{m,n}^\mathrm{(QApwO)}\times
\begin{cases}
\left(\frac{s}{s_0}\right)^{m/2}, & s < s_0,\\
\left(\frac{s}{s_0}\right)^{1/2},       & s \ge s_0,
\end{cases}\\
B_\mathrm{QApwO}(\theta,\zeta) &=& \sum_{m,n} B_{m,n}^\mathrm{(QApwO)} \cos(m\theta -nN_{fp}\zeta)\,,
\end{eqnarray}
with $s_0=0.15$. Then, all the \textit{model} configurations {(model in the sense of not being derived from magnetohydrodynamic equations)} are scaled to a major radius $R=18.5\,$m, and an on-axis magnetic field strength $B_\mathrm{axis}=5.5\,$T (for the model fields, this is done by setting $B_{0,0}^\mathrm{(QApwO)}=B_\mathrm{axis}$ and $B_\zeta = R B_\mathrm{axis}$). 

The neoclassical code \texttt{MONKES} is able to compute the neoclassical transport coefficients~\cite{beidler2011ICNTS}. Convolution (including momentum conservation) of these coefficients to compute the neoclassical radial fluxes and parallel flows for a given set of plasma profiles is provided by the suite \texttt{Neotransp}~\cite{smith2020neotransp} (some calculations were later repeated with \texttt{SFINCS}~\cite{landreman2014sfincs}). We consider a deuterium-tritium plasma with density and temperature \cite{landreman2022bootstrap}
\begin{equation}
n_b(s)=n_{b,0}(1-s^5)\,,\quad T_b(s)=T_{b,0}(1-s)\,,
\end{equation}
with $n_{e,0}=2n_{D,0}=2n_{T,0}$ and $T_{e,0}=T_{D,0}=T_{T,0}$. Once the total bootstrap current is computed \jl{(including the contribution of all species, the effect of the ambipolar radial electric field and momentum conservation)}, the change in $\iota$, shown in Figure~\ref{FIG_I}, is estimated as
\begin{equation}
\Delta\iota(s) = \frac{\mu_0R}{2 s B_\mathrm{axis}^2}\int_0^{s} \fsa{\mathrm{j}_\parallel\cdot\mathrm{B}}(s') \mathrm{d}s'\,,
\end{equation}
where $\mu_0$ is the vacuum permeability.

For figure~\ref{FIG_AMB}, we set $n_{e,0}=4\times 10^{20}\,$m$^{-3}$ and $T_{e,0}=12\,$keV and perform the calculation at $s=s_0$.

\end{document}